\documentclass[letterpaper, 10 pt, conference]{ieeeconf}
\IEEEoverridecommandlockouts                              

\usepackage{times}
\usepackage{macros}
\usepackage[utf8]{inputenc} 
\usepackage[T1]{fontenc} 
\usepackage{todonotes}
\usepackage{hyperref} 
\usepackage{url} 
\usepackage{booktabs} 
\usepackage{amsfonts} 
\usepackage{nicefrac} 
\usepackage{microtype} 
\usepackage{xcolor} 
\usepackage{graphicx}
\usepackage{algorithm}
\usepackage{algorithmicx}
\usepackage{algcompatible}
\usepackage{amsmath}
\usepackage{amssymb}
\usepackage{hyperref}
\usepackage{subcaption}
\usepackage{multirow}
\usepackage{arydshln}
\usepackage[table]{xcolor}
\usepackage{float}
\usepackage{algpseudocode}
\usepackage{cite}

\usepackage[normalem]{ulem}

\title{\LARGE \bf
Orchestrating Level-$K$ Policies Against Unknown Opponents in Partially-Observable Dynamic Games
}

\author{Addison Kalanther$^{1}$, Sanika Bharvirkar$^{1}$, Daniel Bostwick$^{1}$, Chinmay Maheshwari$^{2}$, and Shankar Sastry$^{1}$
\thanks{$^{1}$A. Kalanther, S. Bharvirkar, D. Bostwick, and S. Sastry are with EECS,
University of California, Berkeley, CA, USA
{\tt\small \{addikala, sbharvirkar, daniel.k.bostwick\}@berkeley.edu, sastry@coe.berkeley.edu}}%
\thanks{$^{2}$C. Maheshwari is with ECE, Johns Hopkins University,
        Baltimore, MD, USA
        {\tt\small chinmay\_maheshwari@jhu.edu}}%
}

\begin{document}

\maketitle
\thispagestyle{empty}
\pagestyle{empty}

\begin{abstract}
Level-\(K\) reasoning generates a hierarchy of policies specialized to opponents with different reasoning levels. When an opponent’s level is unknown, a common deployment rule estimates that level and selects the corresponding response. In a dynamic, partially-observed game, this selection is repeated, with each choice shaping subsequent states and observations. The response associated with the most likely opponent level need not maximize expected return from the current history. We formulate this deployment problem as dynamic orchestration of a fixed, pretrained policy library in a partially observable Markov game. We compare classification-based orchestrators (CBOs) trained using offline data or on-policy data aggregation with a reinforcement-learning-based orchestrator (RLBO) trained to maximize expected discounted return. In pursuit–evasion experiments, on-policy training improves classification and return, yet RLBO achieves higher return than the on-policy and offline CBOs. Given a pretrained library, RLBO also reaches performance comparable to a policy trained directly over the pursuers’ action space with fewer training timesteps. These findings support treating a hierarchy of level-\(K\) policies as a resource for orchestration, not a prescription for deployment.
\end{abstract}

\section{Introduction}
Autonomous agents in adversarial settings need to reason about how their opponents make decisions. This raises a basic question: how should an agent model an opponent's reasoning? One approach is bounded rationality, which recognizes that decision makers operate with limited information, computational resources, and time \cite{simon1955behavioral,simon1997models}. Level-$k$ reasoning represents strategic reasoning through a hierarchy of responses to less sophisticated opponents \cite{stahl1995players,nagel1995unraveling}. Together with the related cognitive hierarchy model \cite{camerer2004cognitive}, it offers a structured alternative to assuming equilibrium play and can explain behavior observed in strategic decision experiments.

These models have attracted attention in control and robotics, including autonomous driving, pursuit-evasion, and vehicle coordination \cite{li2018game,li2019decision,tian2021anytime,kokolakis2020bounded,guan2021bounded,kokolakis2023bounded}. In the standard level-$k$ construction, a level-$0$ policy specifies baseline behavior, and a level-$(k+1)$ policy is computed as a best response to a level-$k$ opponent. For dynamic games, these responses specify behavior over an evolving interaction. Computing them offline produces a library of policies specialized to different opponent reasoning levels. This hierarchy specifies how responses are constructed; deciding how to use them during deployment is a separate problem.

This distinction becomes important when the opponent's reasoning level is unknown. A policy optimized against one level may perform poorly against another. In a dynamic game, an agent must act while learning about its opponent, and its actions change the physical state and can affect subsequent observations. Deployment therefore involves repeated policy choices during a partially-observed interaction.

One approach maintains a belief over the opponent's reasoning level using interaction history and plans future actions under that belief. Li et al. combine Bayesian inference with predictive control \cite{li2019decision}, while Tian et al. use a partially-observable decision formulation and belief-space search to reason about latent human states \cite{tian2021anytime}. Although precomputed opponent models and value functions can support such methods, online belief propagation and multi-step action optimization can remain computationally demanding as the state space and planning horizon grow. Selecting from a pretrained response library offers a complementary way to reuse offline computation during deployment.

Another approach estimates the opponent's level online and deploys the corresponding response. Guan et al. use maximum likelihood estimation to adapt the agent's reasoning level during pursuit-evasion \cite{guan2021bounded}. Ho et al. study a related Bayesian rule that samples an opponent level from a belief distribution and responds one level higher \cite{ho2021bayesian}. The pairing of an opponent level with its response thus becomes a rule for deployment. Ho et al. also note that a best response to a distribution over opponent levels can fall outside the level-$k$ hierarchy.

Using this pairing during deployment raises two issues. First, when several opponent levels remain possible, the response paired with the most likely level need not yield the highest expected return. Classification accuracy also ignores how costly different estimation errors are. Second, switching policies mid-trajectory can produce states and histories unlike those encountered during an expert’s training. An approximately-learned best response may then be a poor choice, even when the opponent’s level is correctly identified.

We study this deployment problem as the orchestration of a fixed library of bounded-rational policies. Related ideas include reward-trained expert gating for opponent modeling \cite{he2016opponent} and Bayesian policy reuse \cite{rosman2016bpr}. Our focus is an orchestrator that can select and switch among separately trained level-$k$ responses at each time step of a partially observed encounter. We seek to use the hierarchy as a resource for decision making, allowing policy selection to be learned without enforcing the original level-to-response pairing. This leads to the following question:

\begin{quote}
How should an agent select and switch among pretrained level-$k$ responses as a partially-observable game evolves, to maximize its expected return against an opponent of unknown reasoning level?
\end{quote}

We formulate this problem in a partially-observable Markov decision process and compare classification-based orchestration with reward-maximizing orchestration using the same fixed response library. The classification methods use either offline data or on-policy, DAgger-style data aggregation \cite{ross2011dagger}; the reward-based method learns policy selection through reinforcement learning.

In our pursuit-evasion experiments, reward-based orchestration achieves higher expected return than both classification-based methods. Training the classifier on histories generated by its own policy choices improves identification and return, but does not close the performance gap. Although the learned responses are specialized to different evaders, reward-based orchestration frequently selects a response other than the one paired with the evader's true level. Given a pretrained response library, orchestration also reaches performance comparable to a policy trained directly over the pursuers' action space, with fewer additional training timesteps. These results support treating the level-$k$ hierarchy as a resource for orchestration, not a prescription for deployment.

\section{Problem Formulation}\label{sec:probform}
We study the problem of dynamically selecting among a fixed
library of bounded-rational policies when interacting with
an opponent whose level of rationality is unknown.
We formulate this problem as a two-player partially-observable Markov game (POMG). Throughout, we refer to
the decision-making agent as the \emph{hero} and its
opponent as the \emph{adversary}. Each player may represent
an individual agent or a team of agents acting under a
shared objective.

\subsection{Game Model}

We consider a two-player POMG defined by the tuple \(
\mathcal{M} =
\langle
\mathcal{S},
\mathcal{A},
\bar{\mathcal{A}},
\mathcal{O},
\bar{\mathcal{O}},
\mathcal{P},
\mathcal{Z},
\bar{\mathcal{Z}},
\mathcal{R},
\bar{\mathcal{R}},
\rho,
\gamma
\rangle.\)
Here, $\mathcal{S}$ is the state space;
$\mathcal{A}$ and $\bar{\mathcal{A}}$ are the action
spaces of the \emph{hero} and \emph{adversary}, respectively; and
$\mathcal{O}$ and $\bar{\mathcal{O}}$ are their
respective observation spaces.
The transition kernel
$\mathcal{P}:
\mathcal{S}\times\mathcal{A}\times
\bar{\mathcal{A}}\rightarrow\Delta(\mathcal{S})$
governs the evolution of the system.
The observation kernels
$\mathcal{Z}:\mathcal{S}\rightarrow
\Delta(\mathcal{O})$ and
$\bar{\mathcal{Z}}:\mathcal{S}\rightarrow
\Delta(\bar{\mathcal{O}})$
specify the observations available to each player.
We assume their observations are conditionally independent
given the state.
The reward functions
$\mathcal{R},\bar{\mathcal{R}}:
\mathcal{S}\times\mathcal{A}\times
\bar{\mathcal{A}}\rightarrow\mathbf{R}$
specify the respective objectives of the two players.
Finally, $\rho\in\Delta(\mathcal{S})$ is the initial
state distribution, and $\gamma\in[0,1)$ is a
shared discount factor.
At each time step $t$, the system evolves according to \(
s_{t+1}\sim
\mathcal{P}(\cdot\mid s_t,a_t,\bar a_t),\)
where $s_t\in\mathcal{S}$ is the state and
$a_t\in\mathcal{A}$ and
$\bar a_t\in\bar{\mathcal{A}}$ are the
actions of the hero and adversary, respectively.
The players receive observations \(
o_t\sim\mathcal{Z}(\cdot\mid s_t),
\bar o_t\sim\bar{\mathcal{Z}}(\cdot\mid s_t).\)

Since the state is not necessarily observable, each
player selects actions based on its local history.
Let \(
h_t=(o_0,a_0,\ldots,a_{t-1},o_t)\)
and \(
\bar h_t=
(\bar o_0,\bar a_0,\ldots,\bar a_{t-1},\bar o_t)\)
denote the histories of the hero and adversary,
respectively. Their corresponding history spaces are
denoted by $\mathcal{H}$ and $\bar{\mathcal{H}}$.

Let $\Pi$ and $\bar{\Pi}$ denote the sets of
history-dependent policies of the hero and adversary.
A policy $\pi\in\Pi$ specifies a distribution
$\pi(\cdot\mid h_t)\in\Delta(\mathcal{A})$,
and similarly,
$\bar\pi\in\bar{\Pi}$ specifies
$\bar\pi(\cdot\mid\bar h_t)
\in\Delta(\bar{\mathcal{A}})$.
For a given policy pair $(\pi,\bar\pi)$,
the interaction induces a trajectory distribution
$p(\tau\mid\pi,\bar\pi)$.
For any finite trajectory prefix
$\tau_{0:T}$, including the terminal state
$s_{T+1}$, this distribution factorizes as \(
p(\tau_{0:T}\mid\pi,\bar\pi)
=\rho(s_0)
\prod_{t=0}^{T}
\Big[
\mathcal{Z}(o_t\mid s_t)
\bar{\mathcal{Z}}(\bar o_t\mid s_t)\cdot
\pi(a_t\mid h_t)
\bar\pi(\bar a_t\mid\bar h_t)
\mathcal{P}(s_{t+1}\mid s_t,a_t,\bar a_t)
\Big].\)

The expected discounted returns of the two
players are defined as \(
J(\pi,\bar\pi)
=
\mathbf{E}_{\pi,\bar\pi}
\left[
\sum_{t=0}^{\infty}
\gamma^t
\mathcal{R}(s_t,a_t,\bar a_t)
\right],
\bar J(\pi,\bar\pi)
=
\mathbf{E}_{\pi,\bar\pi}
\left[
\sum_{t=0}^{\infty}
\gamma^t
\bar{\mathcal{R}}(s_t,a_t,\bar a_t)
\right].\)

Each player seeks to maximize these objectives.

\subsection{Bounded-Rational Policy Libraries}
\label{sec:brpl}

We construct the policy libraries of the hero and adversary
using level-$k$ reasoning, a model of bounded rationality
in which agents iteratively best respond to opponents
with lower levels of strategic reasoning.
We begin with a prescribed level-0 policy
for the adversary $\bar{\pi}_0 \in \bar{\Pi}$,
which represents baseline behavior that do not
explicitly account for the hero's strategy.
For each level $k \geq 1$, the corresponding
policies are recursively defined as
\begin{align*}
\pi_k
\in \arg\max_{\pi \in \Pi}
J(\pi,\bar{\pi}_{k-1}),
\label{eq:hero_level_k}
\quad \bar{\pi}_k
\in \arg\max_{\bar{\pi} \in \bar{\Pi}}
\bar{J}(\pi_{k-1},\bar{\pi}).
\end{align*}

Thus, a level-$k$ policy of either player is
constructed as a best response to a level-$(k-1)$
policy of its opponent.

Let $K$ denote the maximum rationality level
considered for the adversary. We construct
the following policy libraries: \(
\mathcal{T}
:= \{\pi_1,\pi_2,\ldots,\pi_{K+1}\},
\bar{\mathcal{T}}
:= \{\bar{\pi}_0,\bar{\pi}_1,\ldots,\bar{\pi}_K\}.\)

By construction, each hero policy
$\pi_{k+1}$ is a best response to the
corresponding adversary policy $\bar{\pi}_k$.

In practice, these policies may be
approximately computed using reinforcement
learning. Throughout this work, we assume
that the policy libraries are generated
offline and remain fixed during deployment.
Our objective is to investigate how the
hero should dynamically select among
these policies when the adversary's
rationality level is unknown.

\subsection{Orchestration Under Unknown Opponent Rationality}
\label{sec:orchestration}

We study the deployment of a fixed library of pretrained
level-$k$ policies,
$\mathcal{T}=\{\pi_1,\ldots,\pi_{K+1}\}$.
Each policy $\pi_{k+1}$ is trained as a best response to
the adversary policy $\bar{\pi}_k$.
At the beginning of an episode, the adversary's level
$L$ is sampled from a known prior
$\mathcal{W}\in\Delta(\{0,\ldots,K\})$.
The adversary follows $\bar{\pi}_L$ throughout the episode,
while $L$ remains unobserved by the hero.

At each history $h_t$, the hero selects an expert to
supply its next action. This choice can be revised at
the next time step. Because the chosen actions affect
the physical state and subsequent observations,
deployment requires a sequence of expert choices
throughout the evolving game.

Formally, an orchestrator is a history-dependent mapping
$\mu:\mathcal{H}\rightarrow\Delta(\{1,\ldots,K+1\})$.
At time $t$, the hero samples
$I_t\sim\mu(\cdot\mid h_t)$ and executes
$a_t\sim\pi_{I_t}(\cdot\mid h_t)$.
The resulting hero policy is
\begin{equation}
\pi_\mu(a\mid h)
=
\sum_{i=1}^{K+1}
\mu(i\mid h)\pi_i(a\mid h).
\label{eq:induced_policy}
\end{equation}
The expert policies remain fixed; only their selection
is controlled by the orchestrator.

Let $\mathfrak{M}$ denote the admissible class of
history-dependent orchestrators. We seek
\begin{equation}
\max_{\mu\in\mathfrak{M}}
\sum_{k=0}^{K}
\mathcal{W}(k)J(\pi_\mu,\bar{\pi}_k),
\label{eq:optimal_orchestration}
\end{equation}
where $J(\pi_\mu,\bar{\pi}_k)$ is the hero's expected
discounted return against $\bar{\pi}_k$.
The prior averages performance over the opponent levels
that may be drawn at the start of an episode, while
history-dependent selection allows adaptation within
the episode.

A common deployment principle estimates the opponent's
level and selects the response one level higher
\cite{guan2021bounded}.
An idealized Bayesian version of this rule uses the posterior
$b_t(k):=\mathbf{P}(L=k\mid h_t)$ to select
\begin{equation}
\hat{k}_t
\in
\arg\max_{k\in\{0,\ldots,K\}}b_t(k),
\qquad
I_t=\hat{k}_t+1,
\label{eq:classification_orchestration}
\end{equation}
with ties resolved by a fixed rule.
In practice, a learned classifier can supply the level
estimate. This strategy is a particular orchestrator
that links expert selection directly to opponent
identification. Two limitations motivate evaluating
this pairing against return-based selection.

\emph{{\color{blue}\textbf{(C1)}} Misalignment between opponent identification and return:}
When several levels remain plausible, an expert that
is best against the most likely level may perform
poorly against the others. A different expert choice
can therefore yield higher expected return, even
when the posterior is exact.
Moreover, classification accuracy counts all incorrect
level estimates equally, although their consequences
for return can differ. Thus, accurate opponent
inference and effective response selection are
distinct objectives.

\emph{{\color{blue}\textbf{(C2)}} Distribution shift under dynamic orchestration:}
Switching among experts can produce histories that
differ from those generated by their original
policy pairings. Even an exact best response given
the initial state distribution need not provide an
optimal continuation at histories outside the support
of its original interaction distribution.
For approximately learned experts, limited training
coverage can also lead to poor performance at
histories reached during orchestration.
Consequently, the paired expert may be a poor choice
even when the opponent's level is correctly identified.
A related shift can affect the opponent classifier:
when trained on fixed policy-pair trajectories, it
may encounter different histories during deployment.
Its predictions influence expert selection and hence
the distribution of histories on which subsequent
predictions are made.

Our formulation does not require opponent identification
as an intermediate objective. The level-$k$ hierarchy
specifies the available responses, while the
orchestrator determines how they are deployed.
We compare classification-based and reward-trained
orchestrators using the same fixed library, with
expected discounted return as the common performance
criterion.

\section{Orchestration Strategies}\label{sec:OrchStrat}
We investigate two approaches to dynamic orchestration
of pretrained level-$k$ policies. The first follows
a classification-based approach, in which the adversary's
rationality level is estimated from the interaction
history and the corresponding best-response policy
is deployed. The second employs reinforcement learning
to directly optimize the selection of pretrained
policies based on their cumulative reward.

For classification-based orchestration, we additionally
investigate two training procedures to examine the
effects of distribution shift between classifier training and deployment.

\subsection{Classification-Based Orchestration}
\label{sec:cbo}

Classification-based orchestration (CBO) estimates the
adversary's rationality level and deploys the
corresponding pretrained best-response policy.

Let \(
q_\psi:\mathcal H \rightarrow
\Delta(\{0,\ldots,K\})\)
denote an opponent classifier parameterized by
$\psi$, where $q_\psi(k\mid h_t)$ represents
the estimated probability that the adversary
has rationality level $k$, given the hero's
interaction history $h_t$.

At each time step, the classifier estimates
the adversary's level according to
\begin{equation}
\hat{k}_t
\in
\arg\max_{k\in\{0,\ldots,K\}}
q_\psi(k\mid h_t).
\label{eq:level_estimation}
\end{equation}
The classification-based orchestrator then
selects the corresponding best-response policy \(
\mu_{q_\psi}(i\mid h_t)
=
\mathbf{1}\{i=\hat{k}_t+1\},\)
where ties in \eqref{eq:level_estimation}
are resolved using a fixed rule.

We consider classifiers trained by minimizing
the cross-entropy loss. Given a dataset
$\mathcal D$ of interaction histories and
their corresponding adversary levels,
the classifier parameters are obtained by
solving
\begin{equation}
\min_\psi
\mathcal L_{\mathcal D}(\psi)
:=
-\mathbf E_{(h,k)\sim\mathcal D}
\left[
\log q_\psi(k\mid h)
\right].
\label{eq:classifier_loss}
\end{equation}

We investigate two procedures for constructing
the training dataset: offline data collection
using fixed policy pairs and on-policy data
aggregation under the classification-based
orchestrator.

\subsubsection{Offline Classification (CBO-Off)}
\label{sec:offline_classifier}

We first consider an offline classifier  trained
on trajectories generated by fixed pairs of
hero and adversary policies. 

At the beginning of each training episode,
a policy pair
$(\pi_i,\bar{\pi}_k)
\in\mathcal T\times\bar{\mathcal T}$
is sampled from a prescribed distribution
$d_{\mathrm{off}}$.
Both policies remain fixed throughout the
episode, and the resulting interaction
histories are labeled with the adversary's
rationality level $k$.

For a given time step $t$, the resulting
joint distribution of histories and
adversary levels is
\begin{equation}
\begin{aligned}
d_{\mathrm{off}}^t(h,k)
&=
\sum_{i=1}^{K+1}
d_{\mathrm{off}}(i,k)\cdot
p(h_t=h\mid\pi_i,\bar{\pi}_k).
\end{aligned}
\label{eq:offline_distribution}
\end{equation}

Unless otherwise specified, we sample
policy pairs uniformly.
The collected history-label pairs form
the offline dataset $\mathcal D_{\mathrm{off}}$,
which is used to train the classifier
according to \eqref{eq:classifier_loss}.

Although the resulting classifier observes
interactions involving different policy
pairs, each training trajectory is generated
using a fixed hero policy. During deployment,
however, the classifier's predictions
determine which policies are executed.
Consequently, the resulting distribution
of interaction histories may differ from
that in offline training.

This motivates our second classification
strategy, which adapts the training data
to the distribution induced by the
classification-based orchestrator itself.

\subsubsection{On-Policy Classification (CBO-On)}
\label{sec:onpolicy_classifier}

To reduce the distribution shift between
classifier training and deployment,
we consider an iterative, DAgger-style
dataset aggregation procedure \cite{ross2011reduction}.

Rather than collecting trajectories
using fixed hero policies, we deploy the
classification-based orchestrator induced
by the current classifier and collect
histories generated by its own
policy-selection decisions.

Specifically, let $q_{\psi_n}$ denote
the classifier at iteration $n$,
and let $\mu_{q_{\psi_n}}$ denote
its corresponding orchestrator.
We sample an adversary level
$L\sim\mathcal W$ and generate a
trajectory by deploying
$\mu_{q_{\psi_n}}$ against
$\bar{\pi}_L$.

The resulting histories are labeled
with the adversary's true rationality
level and aggregated into a training
dataset:
\begin{equation}
\mathcal D_{n+1}
=
\operatorname{Buffer}_C
\left(
\mathcal D_n
\cup
\{(h_t,L)\}_{t=0}^{T_n}
\right),
\label{eq:dagger_dataset}
\end{equation}
where $T_n$ denotes the final time
step of the collected episode and
$\operatorname{Buffer}_C$ is  
a first-in-first-out buffer of
capacity $C$.

The classifier is subsequently
updated by minimizing the
cross-entropy loss over the
aggregated dataset.
The complete training procedure
is summarized in Algorithm~\ref{alg:dagger}.

\begin{figure}[t]
\centering
\begin{minipage}{\columnwidth}
\begin{algorithmic}[1]
\Require
Policy libraries
$\mathcal T,\bar{\mathcal T}$;
opponent distribution $\mathcal W$;
buffer capacity $C$;
training iterations $N_{\mathrm{DAgger}}$;
training epochs $E$.

\Ensure
Trained classifier $q_{\psi}$.

\State Initialize classifier parameters $\psi$
and buffer $\mathcal D\gets\emptyset$.

\For{$n=1,\ldots,N_{\mathrm{DAgger}}$}

    \State Sample adversary level
    $L\sim\mathcal W$.

    \State Initialize a new episode
    against $\bar{\pi}_L$.

    \While{episode has not terminated}

        \State Observe hero history $h_t$.

        \State Estimate opponent level:
        \[
        \hat{k}_t
        \in\arg\max_k q_\psi(k\mid h_t).
        \]

        \State Select expert
        $I_t=\hat{k}_t+1$.

        \State Execute
        $a_t\sim\pi_{I_t}(\cdot\mid h_t)$.

        \State Append $(h_t,L)$ to
        $\mathcal D$, evicting the
        oldest entries if $|\mathcal D|>C$.

    \EndWhile

    \For{$e=1,\ldots,E$}

        \State Sample minibatches
        $(h,k)\sim\mathcal D$.

        \State Update $\psi$ by
        gradient descent on \(
        -\log q_\psi(k\mid h).\)
    \EndFor

\EndFor

\State \Return $q_\psi$.
\end{algorithmic}
\end{minipage}
\caption{On-policy opponent classifier training
using DAgger-style dataset aggregation.}
\label{alg:dagger}
\end{figure}

By repeatedly collecting data under
the current orchestrator, this
procedure exposes the classifier
to histories induced by its own
policy-selection decisions.
This reduces reliance on the
fixed-policy interaction distribution
used for offline classification.

Nevertheless, both classification
strategies optimize the same
opponent-identification objective.
Even when the classifier accurately
estimates the opponent's rationality
level, selecting the corresponding
pretrained best response need not
maximize the hero's expected return.

Moreover, on-policy classifier
adaptation does not directly
address the distribution shifts
encountered by the pretrained
expert policies themselves.

\subsection{Reinforcement-Learning-Based Orchestration }
\label{sec:rl_orchestration}

Classification-based orchestration
uses opponent identification as an
intermediate step for selecting
pretrained policies. We alternatively
consider learning an orchestrator
that directly optimizes the hero's
expected discounted return.

Let \(
\mu_\phi:
\mathcal H
\rightarrow
\Delta(\{1,\ldots,K+1\})\)
denote a history-dependent
orchestrator parameterized by $\phi$.
At each time step $t$,
the orchestrator selects an expert
according to \(
I_t\sim\mu_\phi(\cdot\mid h_t).\)

The selected expert subsequently
determines the hero's action: \(
a_t\sim\pi_{I_t}(\cdot\mid h_t).\)
As defined in
\eqref{eq:induced_policy},
the resulting hero policy is \(
\pi_{\mu_\phi}(a\mid h)
=
\sum_{i=1}^{K+1}
\mu_\phi(i\mid h)
\pi_i(a\mid h).\)

We optimize the orchestrator
parameters by maximizing
the expected discounted return
over the distribution of
adversary levels:
\begin{equation}
\max_\phi
\sum_{k=0}^{K}
\mathcal W(k)
J(\pi_{\mu_\phi},\bar{\pi}_k).
\label{eq:rl_orchestration}
\end{equation}

This objective corresponds to the
general orchestration problem
introduced in
\eqref{eq:optimal_orchestration},
restricted to the parameterized
orchestrator class
$\{\mu_\phi\}$.

We formulate this optimization
as a reinforcement learning problem
in which the orchestrator's
action space consists of the
available pretrained experts.

During training, an adversary
level is sampled from
$\mathcal W$ at the beginning
of each episode and remains
fixed throughout the interaction.
The orchestrator observes the
hero's interaction history,
selects an expert, and receives
the rewards resulting from
the interaction.

The pretrained expert policies
remain fixed, and only the
orchestrator parameters are
updated.

Unlike classification-based
orchestration, this approach
does not explicitly optimize
opponent-identification accuracy.
Instead, the orchestrator learns
policy-selection decisions from
their effects on cumulative reward.

Training the orchestrator on
trajectories generated by its
own policy-selection decisions
also exposes it to the
interaction distributions
induced by dynamic switching.

Thus, reinforcement-learning-based
orchestration (RLBO) provides an
alternative mechanism for
addressing opponent uncertainty
and adapting policy selection
to the consequences of
previously deployed experts.

\section{Case Study on Pursuit-Evasion Games}\label{sec:Case}
We evaluate the orchestrators of Section~\ref{sec:orchestration} in a partially-observable pursuit-evasion game involving a team of two pursuers (the hero) and one evader (the adversary). We design the environment to capture features central to the study of this orchestration problem. The evader's reasoning level must be inferred through interaction and obstacles make visibility intermittent, so policy selection affects both pursuer and evader state, as well as the information available for later decisions.

\subsection{Environment and Information Structure}
\label{sec:environment}

The pursuer team consists of two heterogeneous agents: a
\emph{High-Level Pursuer} (HLP) and a \emph{Low-Level Pursuer} (LLP).
The HLP operates above the buildings and is equipped with a
downward-facing camera, providing a relatively large field of view.
The LLP operates at the same altitude as the evader (EDR) and is
equipped with a forward-facing camera. Its field of view is smaller,
but it can observe regions that are occluded from the HLP. The LLP is
also the pursuer capable of intercepting the EDR.

The two pursuers communicate throughout the episode. In particular,
they share their positions, headings, and any available information
about the current or previously observed location of the EDR. Hence,
although sensing is decentralized, we model the HLP and LLP as a
single pursuer team with a shared observation history and objective.

\begin{figure}[t]
    \centering
    \includegraphics[scale=0.25]{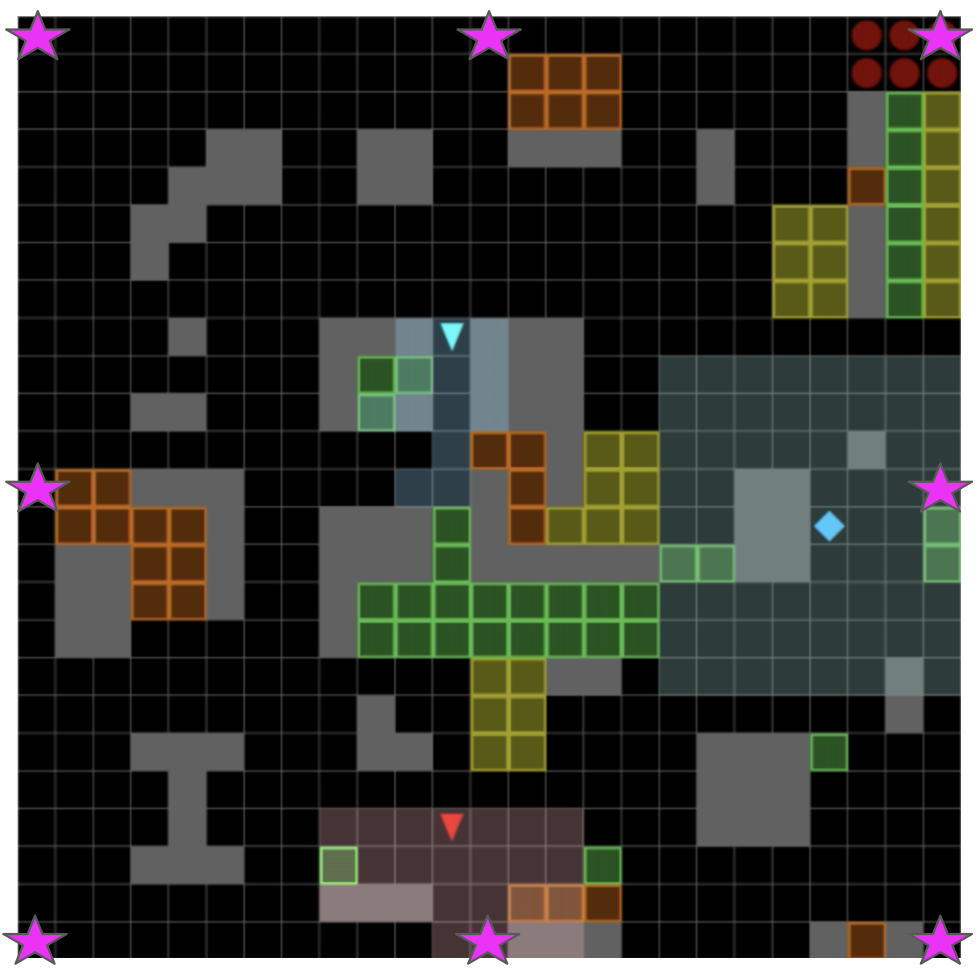}
    \caption{
    Partially observable pursuit--evasion environment.
    Pink stars denote possible evader origin and destination
    locations. The HLP, LLP, and EDR are shown by a blue diamond,
    blue triangle, and red triangle, respectively. Shaded regions
    denote their fields of view. Gray, orange, yellow, and green
    cells correspond to obstacle Types I--IV.
    }
    \label{fig:EnvironmentDescription}
\end{figure}

We model the environment as a two-dimensional grid
$\mathcal{G}=[N]\times[N]$. Obstacles impose both motion and
visibility constraints. Let
$\mathcal{G}_{I},\mathcal{G}_{II},
\mathcal{G}_{III},\mathcal{G}_{IV}\subseteq\mathcal{G}$
denote the four obstacle classes. Their accessibility and visibility
properties are:
\begin{itemize}
    \item \textbf{Type I:} inaccessible and non-viewable to all agents;
    \item \textbf{Type II:} accessible and viewable only to the EDR;
    \item \textbf{Type III:} accessible only to the EDR and viewable
    by the LLP and EDR;
    \item \textbf{Type IV:} accessible and viewable by the LLP and EDR.
\end{itemize}

As such, the feasible position sets of the three agents are \(
    \mathcal{G}_{HLP}
    = \mathcal{G},
    \mathcal{G}_{LLP}
    = \mathcal{G}
    \setminus
    \left(
    \mathcal{G}_{I}
    \cup
    \mathcal{G}_{II}
    \cup
    \mathcal{G}_{III}
    \right),
    \mathcal{G}_{EDR}
    = \mathcal{G}\setminus\mathcal{G}_{I}.\)

At time $t$, let
$\pHLP^t$, $\pLLP^t$, and $\pEDR^t$
denote the positions of the HLP, LLP, and EDR, respectively.
The LLP and EDR additionally have orientations $\orLLP^t,\orEDR^t\in\{0^\circ,90^\circ,180^\circ,270^\circ\}.$
The orientation of the HLP is omitted because its camera is
downward-facing.

Each agent has a time-varying field of view,
denoted by
$\fovHLP^t$, $\fovLLP^t$, and $\fovEDR^t$.
These are determined by the agent positions, orientations, sensing
ranges, and obstacle-induced occlusions.
A grid cell is visible to an agent only if the
line segment connecting the agent to the cell is not intersected by
an obstacle that blocks visibility for that agent. This produces the
occluded sensing regions illustrated in
Figure~\ref{fig:EnvironmentDescription}.

\subsection{POMG Instantiation}
\label{sec:pursuit_pomg}

We instantiate the POMG
introduced in
Section~\ref{sec:probform} by treating the pursuer team as
the hero and the EDR as the adversary. The joint pursuer policy and
evader policy are denoted by \(
    \pi
    =
    \left(
    \pi_{\mathrm{HLP}},
    \pi_{\mathrm{LLP}}
    \right),
    \bar{\pi}
    =
    \pi_{\mathrm{EDR}}.\)
At each timestep, the pursuer team and evader chooses their desired positions and headings, which serves as their actions. 
Each agent is subsequently moved toward its
selected target using an A* path planner, constrained by a set speed. 

The pursuer observation contains the positions and sensing regions
of both pursuers together with the evader state whenever the evader is
visible to either pursuer: \(
o_t = \{
\fovHLP^t,
\fovLLP^t,
\pHLP^t,
\pLLP^t,
\orLLP^t,
(\pEDR^t,\orEDR^t)\) \(
\mathbf{1}
\{
\pEDR^t
\in
\fovHLP^t\cup\fovLLP^t
\}
\}.
\)

Let $g$ denote the EDR's origin--destination pair: $g: (g_\text{origin}, g_\text{destination}) \in \mathcal{G} \times \mathcal{G}$. Unlike the
pursuer team, the EDR observes $g$. Its observation is \(
\bar{o}_t = \{
\fovEDR^t,
\pEDR^t,
\orEDR^t,
g,
\cup_{j\in\{{HLP},{LLP}\}}
(p_j^t,\phi_j^t)
\mathbf{1}
\{
p_j^t\in\fovEDR^t
\}
\}.\)

The pursuer team seeks to intercept the evader before it reaches its
destination, while the evader seeks to reach its destination without
being intercepted.

\paragraph{Reward and termination}
The reward for both agents is a sparse terminal reward, capture, $\mathbf{1}[p^t_{LLP} = p^t_{EDR}]$, for the pursuer team and reaching the destination, $\mathbf{1}[p^t_{EDR} = g_\text{destination}]$, for the evader. The episode terminates either when the evader reaches its destination, is captured by LLP, or max steps has been exceeded.

This partially observable environment is useful for studying dynamic
policy orchestration because different expert-selection decisions can
produce different physical states and observation histories. Hence,
an expert that is effective when deployed from the initial-state
distribution need not remain the best continuation policy after an
arbitrary sequence of previously deployed experts.

\subsection{Bounded-Rational Policy Library}
\label{sec:policy_library_experiment}

We construct the pursuer and evader policy libraries according to the
level-$k$ framework described in Section \ref{sec:brpl}.
Starting from prescribed
level-$0$ policies, successive policies are trained as approximate
best responses to the preceding opponent level using PPO \cite{schulman2017proximal}. This produces the
libraries \(
    \mathcal{T}
    =
    \{\pi_1,\ldots,\pi_{K+1}\},
    \bar{\mathcal{T}}
    =
    \{\bar{\pi}_0,\ldots,\bar{\pi}_K\}.\)

At test time, the evader level
$L\sim\mathcal{W}$ is sampled at the start of an episode and
remains fixed throughout the episode. The pursuer team does not
observe $L$ and must select among pretrained experts
$\pi_i\in\mathcal{T}$ using its interaction history.

\subsection{Orchestration Methods and Baselines}
\label{sec:experimental_methods}

We compare the classification-based and reinforcement-learning-based orchestration methods introduced in Section~\ref{sec:orchestration}, CBO-Off, CBO-On, and RLBO,
together with two additional baselines.

\paragraph{Primitive Action Policy (PAP)}
To determine whether restricting the pursuer to the pretrained expert
library sacrifices substantial performance, we train a primitive
action policy directly against the distribution over evader levels.
The policy satisfies \(
    \pi_{\mathrm{PAP}}(\cdot\mid h)
    \in
    \Delta(\mathcal{A}),\)
and is trained to maximize \(
    \sum_{k=0}^{K}
    \mathcal{W}(k)
    J(
    \pi_{\mathrm{PAP}},
    \bar{\pi}_k
    ).\)
PAP therefore provides an unconstrained primitive-action RL baseline
against which the performance of expert orchestration can be
compared.

\paragraph{Time-Delayed Oracle (TDO).}
Finally, we introduce a time-delayed oracle to study whether revealing
the true opponent level is sufficient to make its corresponding
pretrained best response effective after an arbitrary interaction
prefix.

Let $L$ denote the true evader level and let
$I_t^{\mathrm{CBO}}$ denote the expert selected by a chosen
classification-based orchestrator. For a delay $d$, the TDO selects
\begin{equation}
I_t^{\mathrm{TDO}}
=
\begin{cases}
I_t^{\mathrm{CBO}}, & t<d,\\
L+1, & t\geq d.
\end{cases}
\label{eq:tdo}
\end{equation}
The resulting primitive action is sampled according to \(
    a_t
    \sim
    \pi_{I_t^{\mathrm{TDO}}}
    (\cdot\mid h_t).\)

When $d=0$, the true evader level is known from the beginning of the
episode and the corresponding pretrained best response is deployed
throughout. Increasing $d$ allows the system to evolve under another
orchestrator before revealing the true evader level. Comparing
performance across values of $d$ therefore tests whether the
originally paired best-response expert remains an effective
continuation policy after histories generated by other experts.

\subsection{Analysis}\label{ssec:Analysis}
We evaluate the orchestration methods along three dimensions. First,
we compare their overall performance and training efficiency. Second,
we investigate why classification-based orchestration underperforms,
separating errors due to classifier distribution shift from the more
fundamental limitation of selecting the pretrained best response
associated with the inferred opponent level. Finally, we verify that
the level-$k$ policy library contains specialized and behaviorally
distinct policies, so that policy orchestration is meaningful in the
considered environment.

\subsubsection{Experimental Setup}
Before presenting results, we summarize our experimental configuration.
All experiments use the $25\times25$ pursuit-evasion environment described
in Section~V, with episodes capped at 100 timesteps and discount factor
$\gamma=0.99$. We construct the bounded-rational policy library with
$K{=}8$ using Algorithm~1: each level is obtained by perturbing the
previous one within an $L_2$ ball of radius
$\eta_{\text{pursuer}}=\eta_{\text{evader}}=0.05$ before training against
it, the level-0 evader mixture is taken over $N{=}10$ weights between
$\pi_{A^*}$ and $\pi_{\text{random}}$, and each of the resulting 17
best-response stages (9 pursuer, 8 evader) is trained independently for
$2.0\times10^7$ environment steps with PPO. PAP, RLBO's meta-controller,
and both CBOs are then trained under a matched budget of
$4.0\times10^7$ environment steps, collected in batches of 250 parallel
environments over 100 steps per iteration. All level-$k$, PAP, and RLBO
policies share an LSTM actor-critic with hidden size 64 and a $[64,32,16]$
ReLU head, optimized with Adam ($\text{lr}=3\times10^{-4}$) under standard
PPO hyperparameters ($\lambda_{\text{GAE}}=0.95$, clip $\epsilon=0.2$,
entropy coefficient $5\times10^{-3}$, value coefficient $0.5$, 10 epochs
per update, sub-batch size $50{,}000$); RLBO's meta-controller differs only
in using a categorical head over the $K{+}1{=}9$ pretrained pursuer levels
in place of PAP's continuous action space. CBO-On's classifier shares the
same $[64,32,16]$ ReLU architecture and is trained via DAgger over a FIFO
buffer of $200{,}000$ labeled transitions, with 5 supervised epochs per
rollout batch, to the same $4.0\times10^7$-step budget; CBO-Off instead
trains once over a fixed dataset collected from all $(K{+}1)\times K$
policy pairs (500 environments $\times$ 100 steps each), for 50 epochs.

\subsubsection{Overall Performance}

\begin{figure}
    \centering
    \includegraphics[width=.8\linewidth]{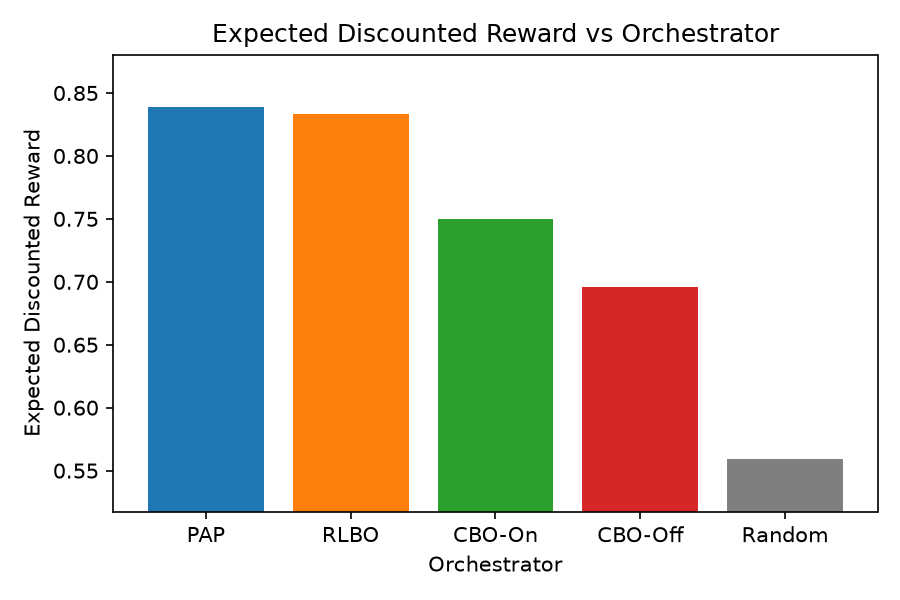}
    \caption{Expected discounted return across orchestration strategies.}
    \label{fig:mpo-vs-cbo}
\end{figure}

Figure~\ref{fig:mpo-vs-cbo} compares the expected discounted return of
RLBO, the offline classification-based orchestrator (CBO-Off), the
on-policy classification-based orchestrator (CBO-On), and random
expert selection. RLBO achieves the highest return among the
orchestration methods, outperforming CBO-On by $11.2$\% and CBO-Off by
$19.8$\%. CBO-On also improves substantially over CBO-Off, suggesting that
adapting the classifier to histories generated by its own
policy-selection decisions is beneficial.

\begin{figure}
    \centering
    \includegraphics[width=.8\linewidth]{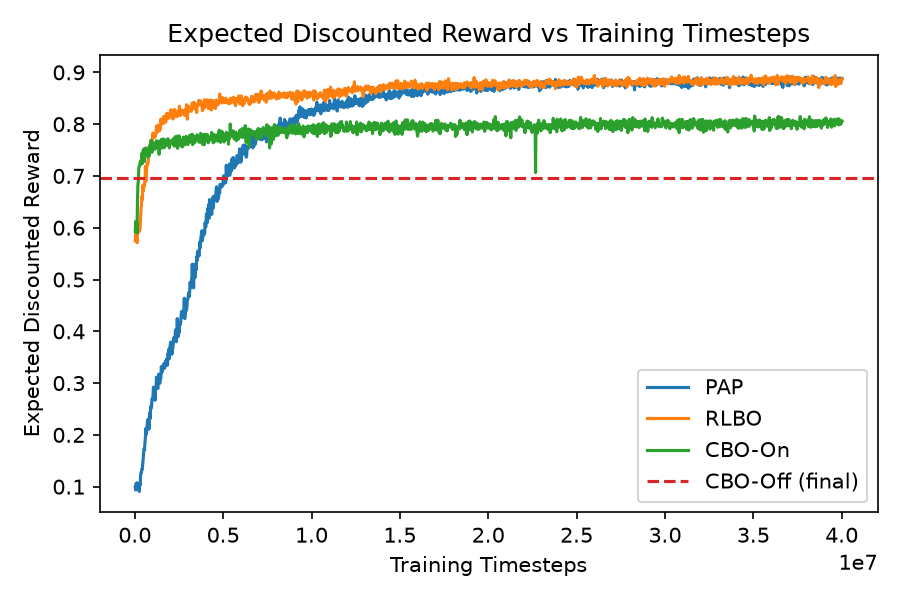}
    \caption{Expected discounted return as a function of orchestration
    training timesteps.}
    \label{fig:rew-vs-ts}
\end{figure}

We compare RLBO with the primitive-action policy (PAP), which
learns directly over the underlying action space rather than selecting
among pretrained experts. As shown in Figure~\ref{fig:rew-vs-ts},
RLBO reaches comparable final performance while requiring $60$\% fewer
training timesteps during orchestration training. 
Although Fig. \ref{fig:rew-vs-ts} omits the sample cost of training level-k policies, the comparison is appropriate, as both methods optimize over the same level-k population, so the recursive computation cost is shared. This suggests that
the pretrained level-$k$ policies provide a useful behavioral prior,
allowing the orchestrator to search over a structured set of
high-level behaviors rather than learning the underlying pursuit
strategy entirely from scratch. 

\subsubsection{Diagnosing Classification-Based Orchestration}

\begin{figure}
    \centering
    \includegraphics[width=.75\linewidth]{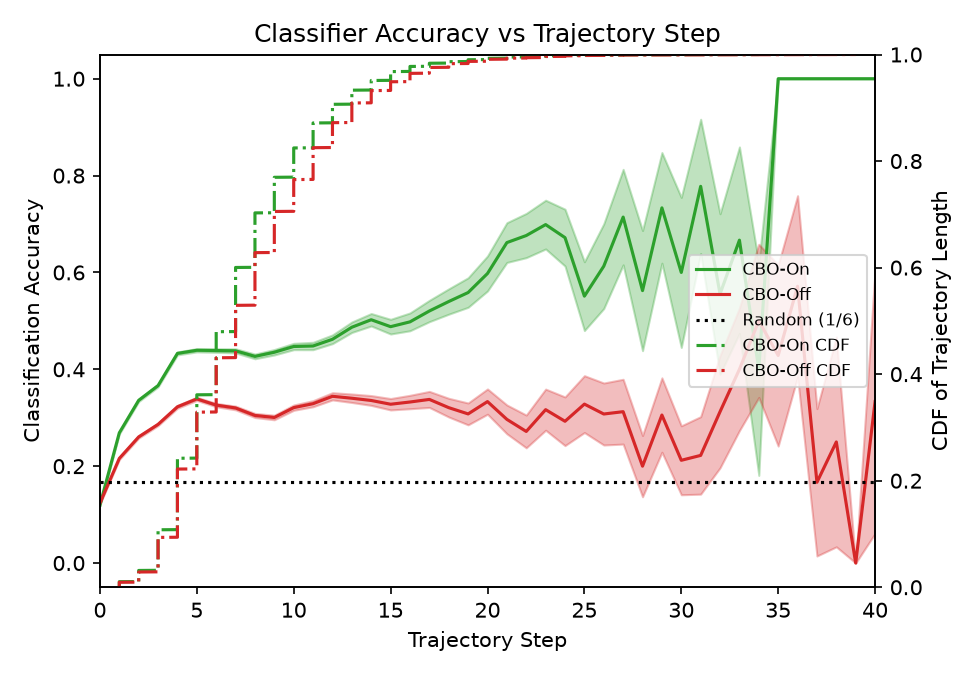}
    \caption{Opponent-level classification accuracy as a function of
    trajectory timestep.}
    \label{fig:classification-perf}
\end{figure}

We first examine whether the performance gap between CBO-Off and
CBO-On can be explained by classifier distribution shift.
Figure~\ref{fig:classification-perf} reports opponent-level
classification accuracy as a function of trajectory length. CBO-On
consistently achieves higher accuracy than CBO-Off. Moreover, its
accuracy generally increases as additional interaction history becomes
available, whereas the accuracy of the offline classifier begins to
decrease after approximately five timesteps.

The difference is consistent with the distribution mismatch induced
by dynamic policy selection. The offline classifier is trained on
histories generated by fixed policy pairs, whereas CBO-On is trained
on histories generated by its own policy-selection decisions. The
improved classification accuracy of CBO-On, together with its improved
return in Figure~\ref{fig:mpo-vs-cbo}, shows that correcting this
mismatch improves classification-based orchestration.

However, improved opponent identification does not close the
performance gap with RLBO. We therefore ask whether the remaining gap
comes from a more fundamental assumption of CBO: that identifying the
opponent level should be followed by deploying the pretrained policy
originally constructed as its best response, a challenge recognized by {\color{blue} \textbf{(C1)}}.

\begin{figure}
    \centering
    \includegraphics[width=.8\linewidth]{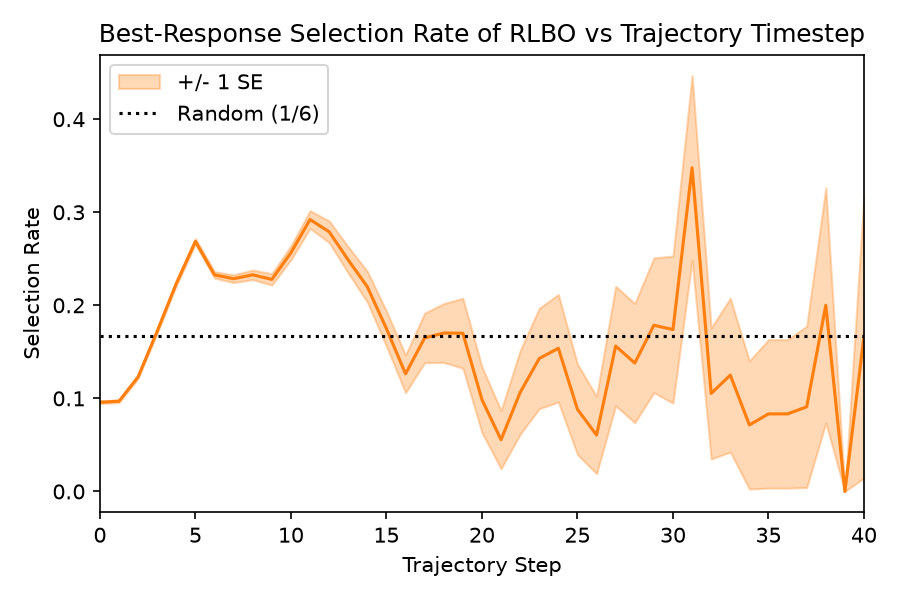}
    \caption{Frequency with which RLBO selects the expert originally
    trained as a best response to the true opponent level. The dashed
    line denotes uniform random expert selection.}
    \label{fig:mpo-classification}
\end{figure}

Figure~\ref{fig:mpo-classification} reports how frequently RLBO selects
the expert paired with the true opponent level. Despite achieving the
highest orchestration return, RLBO selects this paired expert only
intermittently and, at later timesteps, at a rate comparable to or
below uniform random selection.

This observation does not by itself imply that the pretrained
best-response policies are ineffective. Each expert is trained as a
best response when deployed from the initial-state distribution.
Dynamic orchestration, however, may bring the system to histories and states outside the distribution seen in original expert training. The
expert paired with the true opponent level therefore need not provide
the best continuation behavior from such histories, a challenge recognized by {\color{blue} \textbf{(C2)}}. 

\begin{figure}
    \centering
    \includegraphics[width=.8\linewidth]{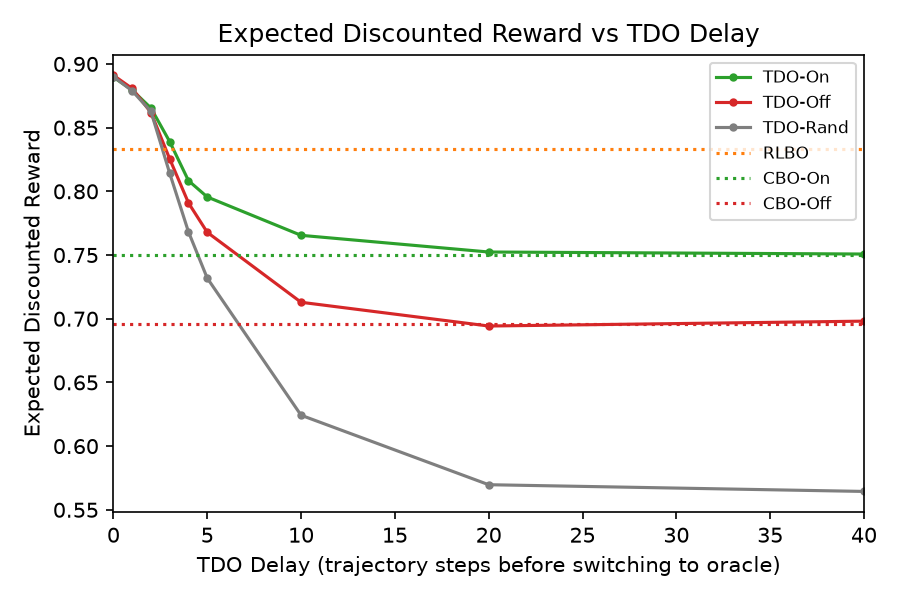}
    \caption{Expected discounted return of the Time-Delayed Oracle as
    a function of the opponent-level reveal time.}
    \label{fig:cbo-vs-tdo}
\end{figure}

We test this hypothesis using the Time-Delayed Oracle (TDO).
For the first $d$ timesteps, the system evolves according to a
non-oracle orchestration strategy. At timestep $d$, the true opponent
level is revealed and the corresponding pretrained best-response
expert is deployed for the remainder of the episode.
Figure~\ref{fig:cbo-vs-tdo} shows that TDO performance deteriorates
rapidly as the reveal time $d$ increases. For sufficiently large
delays, even perfect knowledge of the opponent level followed by its
paired best response performs below RLBO.

Together, Figures~\ref{fig:mpo-classification} and
\ref{fig:cbo-vs-tdo} suggest that the limitation of
classification-based orchestration is not solely opponent
misclassification. The correspondence between an opponent level and
its pretrained best response is defined with respect to deployment
from the initial-state distribution. After a history generated by
other experts, this correspondence may no longer identify the expert
with the highest continuation value. RLBO avoids imposing this
level-to-expert mapping and instead learns expert-selection decisions
directly from their effect on cumulative reward.

\subsubsection{Expert Specialization and Diversity}

\begin{figure}
    \centering
    \includegraphics[width=.6\linewidth]{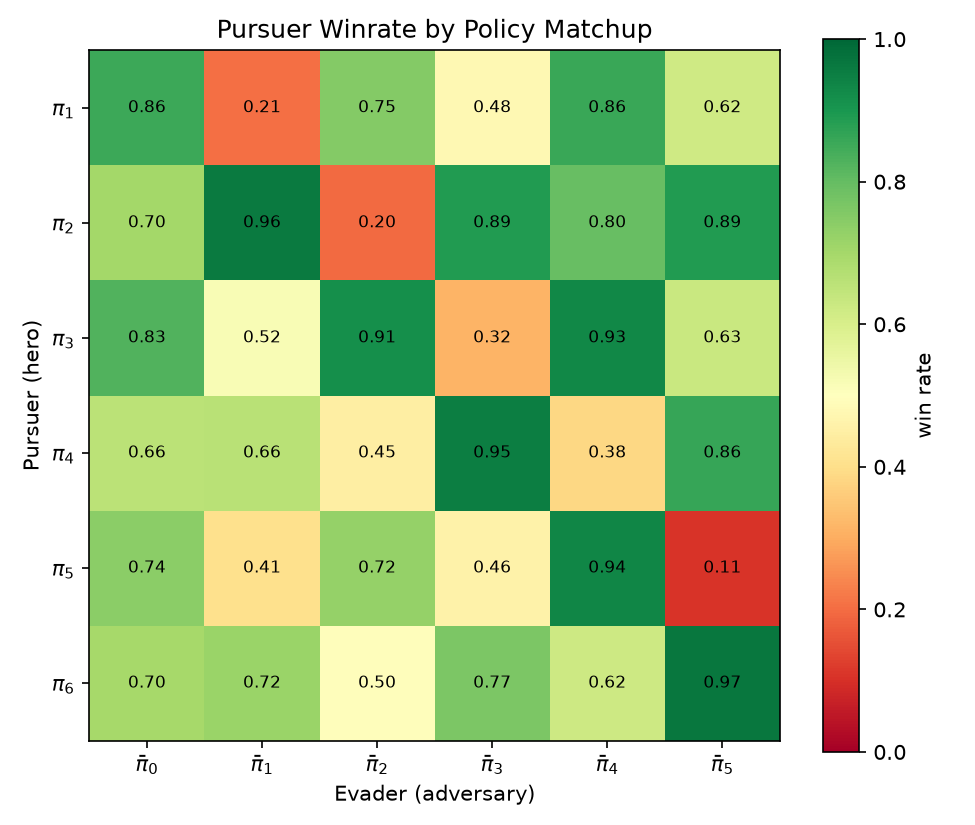}
    \caption{Pursuer win rates for different pursuer--evader policy
    pairs.}
    \label{fig:winrates}
\end{figure}

Finally, we verify that the performance differences above are not an
artifact of an uninformative expert library. We first examine the
pairwise performance of the level-$k$ policies when deployed from the
beginning of an episode. Figure~\ref{fig:winrates} shows substantial
variation across pursuer--evader policy pairs. In particular, the
experts constructed as approximate best responses achieve higher
win rates against their corresponding opponent policies than the
alternative experts, with an average improvement of 32\%.

Thus, the level-$k$ construction produces meaningful specialization:
the identity of the deployed pursuer policy matters when responding
to a fixed evader policy.

We quantify behavioral diversity through the pairwise
total variation distance (TVD) between the pursuer policies' action
distributions. For histories encountered under RLBO, CBO-On,
CBO-Off, and random orchestration, we find the mean pairwise TVD is $0.55$ with a standard deviation of $0.024$, indicating that the experts induce different action distributions on the histories encountered during orchestration.

Together, the win-rate and TVD analyses show that the policy
library contains specialized and distinct experts. The performance advantage of RLBO therefore arises in a setting where selecting among experts is consequential, rather than from a library of similar, interchangeable policies.

\section{Conclusion}
We study bounded-rational policy orchestration in partially-observable games. We present classification-based orchestration (CBO) and analyze its conceptual challenges with distribution shifts. We then propose reinforcement-learning-based orchestration (RLBO) to overcome such challenges in the tested setting. We formulate a pursuit-evasion case study to empirically study their differences. We show RLBO outperforms CBO due to these challenges. While performant, RLBO is more difficult to train. Thus, future work will explore efficient orchestration and performance bounds.

\bibliographystyle{IEEEtran}
\bibliography{refs}

\end{document}